# UNBIASED RATIO-TYPE ESTIMATOR USING TRANSFORMED AUXILIARY VARIABLE IN NEGATIVE CORRELATION CASE

JAYANT SINGH

Department of Statistics, Rajasthan University, Jaipur, India

H.P. Singh

School of Studies in Statistics, Vikram University, Ujjain(M.P.)

†RAJESH SINGH

Department of Statistics, BHU, Varanasi(U.P.), India

†Corresponding author

**Abstract**

The objective of this paper is to propose an unbiased ratio-type estimator for finite population mean when the variables are negatively correlated. Hartley and Ross[2] and Singh and Singh [6] estimators are identified as particular cases of the proposed unbiased estimator. The variance expression of the proposed estimator to the first degree of approximation has been obtained. An empirical study is carried out to demonstrate the performance of the proposed estimator over $\bar{y}$, Robson [5] estimator and Singh and Singh [6] estimator.

*Key words* -  Study variable, auxiliary variate, finite population mean, unbiased estimator, variance.

## *1. Introduction*

It is common practice to use the auxiliary variable for improving the precision of the estimate of a parameter. Out of many ratio and product methods of estimation are good illustrations in this context. When the correlation between the study variate and the auxiliary variate is positive (high) ratio method of estimation is quite effective. On the other hand, when this correlation is negative (high) product method of estimation can be employed effectively.

In fixed population approach, it is well established that the ratio method of estimation does not provide exact expressions for the bias and variance while product method of estimation gives the exact expressions. Considering this fact Srivenkataramana and Tracy [8,9] suggested a transformation for auxiliary variable which enables us to use the product method of estimation in positive correlation case. However, we have used this transformation for defining the ratio-type estimator in negative correlation situation. It would be worth mentioning that the transformation suggested by Srivenkataramana and Tracy [8,9] is a generalization of Srivenkataramana [7], and Bandyopadhyay [1].

Consider a finite population $U = (U_1, U_2, \ldots, U_n)$ of size N. Let (y,x) be the study and auxiliary variates respectively. Suppose $(y_i, x_i)$, (i= 1, 2, …, n) denotes a simple random sample of n observations on characteristic y and x drawn without replacement from U. Further, let $(\bar{y}, \bar{x})$ denote the sample mean estimators of $(\bar{Y}, \bar{X})$ the population means of (y,x) respectively. We assume that x is negatively correlated with y and the population mean $\bar{X}$ of x is known. In this situation, for estimating the population mean $\bar{Y}$, Robson [5] suggested the use of product estimator

$$d_1 = \bar{y}(\bar{x}/\bar{X}) \qquad (1.1)$$

which is rediscovered by Murthy [3].

Using the transformation

$$x_i^* = (N\bar{X} - nx_i)/(N-n), \text{ (i=1,2,…,N)},$$

Srivenkataramana [7] and Bandyopadhyay [1] proposed a dual to product estimator for $\bar{Y}$ as

$$d_2 = \bar{y}(\bar{X}/\bar{x}^*) \qquad (1.2)$$

where $\bar{x}^* = (N\bar{X} - n\bar{x})/(N - n)$.

It is well known that both the estimator's $d_1$ and $d_2$ are biased. In some situations bias becomes disadvantageous and to get rid of this Robson[5] suggested an unbiased product-type estimator

$$d_1^{(u)} = \frac{n(N-1)}{N(n-1)} \bar{y}\left(\frac{\bar{x}}{\bar{X}}\right) - \frac{(N-n)}{N(n-1)} \frac{\bar{p}}{\bar{X}} \tag{1.3}$$

where $\bar{p} = \sum_{i=1}^{n} p_i / n$, $p_i = y_i x_i$, (i=1,2,…,n).

Recently, following the same approach as adopted by Hartley and Ross [2], Singh and Singh [6] suggested an unbiased version of $d_2$ as

$$d_2^{(u)} = \bar{r}^*\bar{X} + \frac{(N-1)}{N(n-1)}\left(\bar{y} - \bar{r}^*\bar{x}^*\right) \tag{1.4}$$

where $\bar{r}^* = \sum_{i=1}^{n} r_i^* / n$, $r_i^* = y_i / x_i$.

The variance of $d_1^{(u)}$ and $d_2^{(u)}$ to the first degree of approximation (or to order $o(n^{-1})$, are respectively given by

$$V(d_1^{(u)}) = \frac{(1-f)}{n}\left[S_y^2 + R^2 S_x^2 + 2R\rho S_y S_x\right] \tag{1.5}$$

$$V(d_2^{(u)}) = \frac{(1-f)}{n}\left[S_y^2 + \bar{R}^{*2} g^2 S_x^2 + 2\bar{R}^* g \rho S_y S_x\right] \tag{1.6}$$

where $S_x^2 = \sum_{i=1}^{N}(x_i - \bar{X})^2/(N-1)$, $S_y^2 = \sum_{i=1}^{N}(y_i - \bar{Y})^2/(N-1)$, $\rho = S_{xy}/(S_x S_y)$

$S_{xy} = \sum_{i=1}^{N}(x_i - \bar{X})(y_i - \bar{Y})/(N-1)$, $R = \bar{Y}/\bar{X}$, $\bar{R}^* = \sum_{i=1}^{N} r_i^* / N$,

g = n/(N-n), f=n/N.

In this paper we have suggested an unbiased ratio-type estimator using the transformation suggested by Srivenkataramana and Tracy [8,9], in the situation where the two variables x and y is negatively correlated and analyzed its properties.

## 2. The Suggested estimator

Using the transformation $u_i = L - x_i$, (i=1,2,…,N), (L, being a scalar) suggested by Srivenkataramana and Tracy [8,9], we define the following ratio estimator for $\bar{Y}$ :

$$d^* = \bar{y}(\bar{U}/\bar{u}) \tag{2.1}$$

where $\bar{u} = \sum_{i=1}^{n} u_i / n$ such that $E(\bar{u}) = \bar{U} = L - \bar{X}$.

Using the standard technique given in Sukhatme and Sukhatme [10], we obtain the bias and variance of $d^*$, to the first degree of approximation, as

$$B(d^*) = \frac{(1-f)}{n} \bar{Y} C_x^2 (\theta + K) \tag{2.2}$$

$$V(d^*) = \frac{(1-f)}{n}\left(S_y^2 + \theta^2 R^2 S_x^2 + 2\theta R \rho S_y S_x\right) \tag{2.3}$$

where $\theta = \bar{X}/(L - \bar{X})$, $C_x = S_x/\bar{X}$ and $K = \rho(C_y/C_x)$.

It is clear from (2.2) that $d^*$ is biased. As our objective is to define unbiased ratio-type estimator for $\overline{Y}$, therefore we will follow the Hartley-Ross [2] procedure for defining an unbiased ratio-type estimator.

Consider a ratio-type estimator
$$d = \overline{v}\ \overline{U} \tag{2.4}$$

where $\overline{v} = \sum_{i=1}^{n} v_i / n$ such that $E(\overline{v}) = \overline{V} = \sum_{i=1}^{N} v_i / N$

The bias of d is given by

$$\begin{aligned} B(d) &= \overline{U}E(\overline{v}) - \overline{Y} = \overline{U}\,\overline{V} - \overline{Y} \\ &= E(u_i) E(v_i) - E(y_i) \\ &= -[E(u_i v_i) - E(u_i) E(v_i)] \\ &= -\text{Cov}(u_i, v_i) \\ &= -\left(\frac{N-1}{N}\right) S_{uv} \end{aligned} \tag{2.5}$$

where $S_{uv} = \sum_{i=1}^{N} (u_i - \overline{u})(v_i - \overline{v})/(N-1)$.

It is well established that

$$\begin{aligned} s_{uv} &= \sum_{i=1}^{n} (u_i - \overline{u})(v_i - \overline{v})/(n-1) \\ &= \frac{n}{(n-1)} (\overline{y} - \overline{u}\,\overline{v}) \end{aligned}$$

is an unbiased estimator of $S_{uv}$ i.e. $E(s_{uv}) = S_{uv}$.

Substitution of $s_{uv}$ in place of $S_{uv}$ in (2.5) yields the unbiased estimator of bias of d as

$$\hat{B}(d) = -\frac{(N-1)n}{N(n-1)} (\overline{y} - \overline{u}\,\overline{v}) \tag{2.6}$$

Hence an unbiased estimator of population mean $\overline{Y}$ is given by

$$d^{(u)} = \overline{v}\,\overline{U} + \frac{(N-1)n}{N(n-1)} (\overline{y} - \overline{u}\,\overline{v}). \tag{2.7}$$

which is Hartley-Ross[2] type estimator.

**Remark 2.1** – For $L = \dfrac{\overline{X}(1+g)}{g}$, $g = n/(N-n)$, $d^{(u)}$ reduces to Singh and Singh [6] estimator

$$d_2^{(u)} = \overline{r}^*\overline{X} + \frac{(N-1)n}{N(n-1)}(\overline{y} - \overline{r}^*\overline{x}^*)$$

while for $L = 0$, it reduces to Hartley-Ross [2] estimator

$$d_3^{(u)} = \overline{r}\overline{X} + \frac{(N-n)n}{N(n-1)}(\overline{y} - \overline{r}\overline{x})$$

which is suitable for the situation where y and x is positively correlated.

Thus we conclude that this study generalizes the work of Hartley-Ross [2] and Singh and Singh [6].

## 3. *Efficiency comparisons*

Following Singh and Singh [6], the variance of $d^{(u)}$ to the first degree of approximation, is given by

$$V(d^{(u)}) = \frac{(1-f)}{n}\left[S_y^2 + \bar{V}S_x^2 + 2\bar{V}\rho S_y S_x\right] \tag{3.1}$$

which is minimum when $\quad \bar{V} = -\beta \tag{3.2}$

where $\beta = \rho(S_y/S_x)$ is the population regression coefficient of y on x.

Under (3.2), the minimum variance of $d^{(u)}$ is given by

$$\min.V(d^{(u)}) = \frac{(1-f)}{n}S_y^2(1-\rho^2) \tag{3.3}$$

It follows from (3.1) and (3.3) that

$$V(d^{(u)}) - \min.V(d^{(u)}) = \frac{(1-f)}{n}(\bar{V}S_x + \rho S_y)^2 \tag{3.4}$$

$$> 0, \text{ unless } \bar{V} = -\beta.$$

Which shows that the minimum variance of $d^{(u)}$ is always less than the variance $d^{(u)}$ unless $\bar{V} = -\beta$.

It can, further be proved that $d^{(u)}$ under $\bar{V} = -\beta$ is more efficient than conventional unbiased estimator $\bar{y}$, Robson [5] estimator $d_1^{(u)}$ and Singh and Singh [6] estimator $d_2^{(u)}$.

Now, we shall make the comparisons of $\bar{y}$, $d_1^{(u)}$, $d_2^{(u)}$ with $d^{(u)}$ when $\bar{V}$ does not coincide with its optimum value $-\beta$.

It is well known under simple random sampling without replacement (SRSWOR) scheme that

$$V(\bar{y}) = \frac{(1-f)}{n}S_y^2 \tag{3.5}$$

We note from (3.1) and (3.5) that $d^{(u)}$ would be better than $\bar{y}$ if

$$\beta < -\frac{\bar{V}}{2} \tag{3.6}$$

It follows from (1.5) and (3.1) that the variance of $d^{(u)}$ is smaller than Robson [5] estimator $d_1^{(u)}$ if

$$\text{Either} \quad \beta < -\frac{R}{2}\left(1+\frac{\bar{V}}{R}\right), \quad \bar{V} > R \tag{3.7}$$

$$\text{Or} \quad \beta > -\frac{R}{2}\left(1+\frac{\bar{V}}{R}\right), \quad \bar{V} < R \tag{3.8}$$

Thus combining the inequalities (3.6) and (3.8) we establish the following theorem

**Theorem 3.1 :** The estimator $d^{(u)}$ is more efficient than $\bar{y}$ or $d_1^{(u)}$ if

$$-\frac{R}{2}\left(1+\frac{\bar{V}}{R}\right) < \beta < -\frac{\bar{V}}{2} \tag{3.9}$$

Further from (1.6) and (3.1) the following theorem can easily be proved.

**Theorem 3.2 :** The estimator $d^{(u)}$ is more precise than $d_2^{(u)}$ if

$$\beta < -\frac{\bar{R}^*}{2}(g + \frac{\bar{V}}{\bar{R}^*}), \quad g < \bar{V}/\bar{R}^* \tag{3.10}$$

It is to be noted that the condition (3.10) is sufficient for $d^{(u)}$ to be better than $\bar{y}$ and $d_2^{(u)}$.

Now form (2.3) and (3.1) we state the following theorem.

**Theorem 3.3 :** The unbiased estimator $d^{(u)}$ would be more efficient than biased estimator $d^*$ if

$$(\theta R + \beta)^2 > (\bar{V} + \beta)^2 \tag{3.11}$$

Thus we infer from (3.11) that $d^{(u)}$ would be more efficient than $d^*$ when $|\beta|$ is closer to $\bar{V}$ than to $\theta R$.

## 4. Empirical study

To see the relative performance of the suggested unbiased estimator $d^{(u)}$ over conventional unbiased estimator $\bar{y}$, Robson [5] estimator $d_1^{(u)}$ and Singh and Singh [6] estimator $d_2^{(u)}$, we use the same population data earlier used by Rao [4]. In this population,

x : the female literacy rate
y : the female work participation rate
N=4, n=2

The required parameters are

$\bar{Y} = 4.87$, $\quad \bar{X} = 43.9175$, $\quad S_x^2 = 31.8575$, $\quad S_y^2 = 4.3118$,

$\bar{R}^* = 0.3099$, $\quad R = 0.1109$, and $\rho = -0.7036$.

We have computed the relative efficiency (%) of $d^{(u)}$ with respect to C for different values of L and compiled in Table 4.1. The relative efficiencies (%) of $d_1^{(u)}$ and $d_2^{(u)}$ with respect to $\bar{y}$ are also given.

**Table 4.1:** Showing the relative efficiencies (%) of $d^{(u)}$ with respect to $\bar{y}$ for different values of L.

| L | 54.70 | 56 | 56.70 | 60.00 | 61.00 | 61.50 | 62.00 |
|---|---|---|---|---|---|---|---|
| RE($d^{(u)}$, $\bar{y}$) | 100.06 | 135.53 | 151.13 | 191.69 | 195.86 | 197.04 | 197.74 |

| L | 62.50 | 63 | 64.50 | 65.00 | 66.50 | 67.00 | 70.00 |
|---|---|---|---|---|---|---|---|
| RE($d^{(u)}$, $\bar{y}$) | 198.02 | 197.97 | 196.31 | 195.40 | 192.04 | 190.79 | 182.81 |

| L | 80.00 | 85.00 | 86.50 | 100.00 | 300.00 | 500.00 | 372607.00 |
|---|---|---|---|---|---|---|---|
| RE($d^{(u)}$, $\bar{y}$) | 160.44 | 152.63 | 150.63 | 137.50 | 107.50 | 104.15 | 100.00 |

RE($d_1^{(u)}$, $\bar{y}$) = 150.00
RE($d_2^{(u)}$, $\bar{y}$) = 191.00

Table 4.1 exhibits that when L ranges :
(i) between 54.70 and 372607.60, the unbiased estimator $d^{(u)}$ is better than simple mean estimator $\bar{y}$.
(ii) Between 56.70 and 86.50, the unbiased estimator $d^{(u)}$ is to be preferred over Robson's [5] estimator $d_1^{(u)}$.
(iii) Between 60 and 67, the unbiased estimator $d^{(u)}$ is more efficient than $\bar{y}$, Robson's [5] estimator $d_1^{(u)}$ and Singh and Singh's [6] estimator $d_2^{(u)}$.
(iv) The maximum efficiency(198.02) is observed when L coincides with 62.50.
Thus we conclude that there is enough scope of choosing L for which the estimator $d^{(u)}$ is better than $\bar{y}$, $d_1^{(u)}$ and $d_2^{(u)}$.